\documentstyle[prb,twocolumn,aps,amsmath,epsfig]{revtex}
\begin{document}
\twocolumn[\hsize\textwidth\columnwidth\hsize\csname@twocolumnfalse%
\endcsname
\title{Electric-field-induced Mott insulating states in organic field-effect transistors}
\author{Olivier C\'epas and Ross H. McKenzie}
\address{Department of Physics, University of
Queensland, Brisbane, 4072, Australia}

\date{\today}
\maketitle

\begin{abstract}
We consider the possibility
that the electrons injected into organic field-effect
transistors are strongly correlated.  A single layer
of acenes can be modelled by
a Hubbard Hamiltonian similar to that used for the $\kappa$-(BEDT-TTF)$_2$X
family of organic superconductors.  The injected electrons do not
necessarily undergo a transition to a Mott insulator state as they
would in bulk crystals when the system is half-filled.  We
calculate the fillings needed for obtaining insulating states in the
framework of the slave-boson theory and in the limit of large Hubbard
repulsion, $U$.  We also suggest that these Mott states are unstable above
some critical interlayer coupling or long-range Coulomb interaction.
\end{abstract}

]

\section{Introduction}

Recent advances in
producing organic field-effect transistors\cite{katz}
and
the growth of single crystals of acenes\cite{Kloc,Anthony}
raises the possibility of           using an electric field
 to inject charge into
a single layer of a layered molecular crystal such as pentacene.
Recent theoretical papers have considered the possibility
of superconductivity due to the electron-phonon
 interaction.\cite{Kato,MacDonald}
Superconductivity might
appear when the filling of the layer at the interface between the
molecular crystal and the dielectric reaches
half-filling\cite{MacDonald}. In that theory, this would be the
consequence of the existence of a van Hove singularity in the density
of states and the pairing is described by the conventional BCS theory
involving the electron-phonon interaction.\cite{MacDonald} However, in
other organic materials, such as the $\kappa$-(BEDT-TTF)$_2$X family,
the proximity of the superconducting phase with various other ground
states (a Mott insulating state for example\cite{Jerome}) may
suggest a more exotic type of pairing, involving the Coulomb
interaction between the electrons.\cite{Kivelson,McKenzie} It is
therefore important to study whether the injected carriers could form
a strongly correlated state (at half-filling for example).

The first question we have to study is whether a Mott insulating state
could be induced.  Naively, one might expect that if the repulsion
between two electrons on the same molecule is strong enough an
insulating state will occur at half filling.  The effect of disorder
on such an insulating state in C$_{60}$ crystals doped by charge
injection was recently considered theoretically.\cite{Sawatzky} Here,
even though we consider a clean system, it is not obvious that the
effects of correlation will appear in a field effect geometry.

In a field-effect transistor, the charges are confined near the
interface between the molecular crystal and the dielectric, because of
the strong electric field used to inject the charges. The issue of how
far extended from the interface the electronic wave-functions are is
very important. As the doping increases, it is indeed not clear
whether the electrons will remain confined to the first layer or because
of the Coulomb interaction the electrons will occupy other states extended far
from the interface.  There is clearly a competition between the
Coulomb repulsion that pushes the electrons apart and the electric
field that forces the electrons to stay close to the interface. Then,
if the electric field is not strong enough, the Mott transition will
not occur. In the following, we will adopt the slave-boson approach to
study the Mott transition in the presence of the electric
field. Slave-boson theory is the simplest theory which generically
gives a Mott transition, and it will allow us to show that such Mott
states could occur at non half-integer fillings.

In section \ref{A strongly}, we will estimate the Coulomb interaction
for two electrons on the same acene molecule, such as
pentacene, tetracene and anthracene since these     are important
examples. We will then give a brief review of the physics of a single
layer of such molecules. However, the presence of a finite electric
field in field-effect transistors allows the electrons to hop onto the
neighbouring layers, thus modifying the dynamics. In section \ref{A
model}, by applying the slave-boson theory, we will show that the Mott
state could occur at larger fillings, that we calculate in the limit
of strong interaction.  In sections \ref{effectof} and
\ref{Inter-site}, we give a discussion of the effect of the
interactions we neglected first, such as the interlayer hopping, and
the long-range Coulomb interactions. Those interactions could in fact
destroy the Mott state, and we will calculate the critical strength
above which the Mott state is no longer stable.

\section{A strongly correlated model for a single layer in acene molecular crystals}
\label{A strongly}

In this section, we consider a model for a layer of acene molecules.
Band structure calculations have shown that, for large electric
fields applied to pentacene\cite{MacDonald} or 
C$_{60}$ (Ref. \onlinecite{Poilblanc})
 field-effect transistors, the electrons are
confined to a single layer.
 
\begin{figure}[tbp]
\centerline{\psfig{file=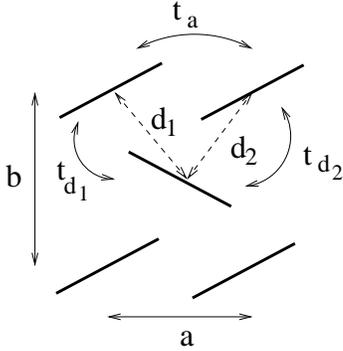,width=4.5cm}}
\vspace{0.2cm}
\caption{Model for a layer of pentacene molecules, according to
Ref. \protect \onlinecite{Cornil,Cornil2}.
 The thick lines represent the pentacene
molecules. $t_a$, $t_{d_1}$ and $t_{d_2}$ denotes the different
hopping integrals between neighboring molecules.}
\label{crystal}
\end{figure}

For acene-based materials, we now argue that a single layer can be
described by a Hubbard model on an anisotropic triangular
lattice. Such a short-range Coulomb interaction is more valid in the
strong doping regime where the screening of the Coulomb interaction by
the other electrons is more efficient. This is the regime we are
interested in. The Hamiltonian is written:

\begin{equation}
{\cal H} = -\sum_{{\bf{ij}},\sigma} t_{\bf ij} (c^{\dagger}_{{\bf{i}},\sigma}
c_{{\bf{j}},\sigma} + \mbox{h.c.})  + U \sum_{{\bf{i}}} n_{{\bf{i}},\uparrow}
n_{{\bf{i}},\downarrow} 
\label{hamiltonian}
\end{equation}
where $c^\dagger_{\bf i \sigma}$ creates an electron [hole] at site
${\bf i}$ in the lowest unoccupied molecular orbital (LUMO) [highest
occupied molecular orbital (HOMO)]. $t_{\bf ij} $ are the
tight-binding hopping integrals between molecular orbitals ($t_{d_1}$,
$t_{d_2}$, $t_a$), as shown in Fig. \ref{crystal}, and $U$ is the
Hubbard repulsion for two electrons on the same molecule that we will
estimate below.  The kinetic part of the Hamiltonian can be rewritten:

\begin{eqnarray}
{\cal H}_0 &=& \sum_{{\bf k},\sigma} \epsilon_k c^{\dagger}_{{\bf
k} \sigma}c_{{\bf k} \sigma} \nonumber \\ \epsilon_k &=& -
2t_{d_1} \cos(k_1 d_1) -2t_{d_2} \cos(k_2 d_2) \nonumber \\ &-& 2t_a
\cos(k_1 d_1 + k_2d_2)
\label{bandstru}
\end{eqnarray}
where $t_{d_{1,2}}$, $t_a$ are the hopping parameters in the $d_{1,2},
a$ directions. For example, $t_{d_1}=59$ meV, $t_{d_2}=88$ meV, $t_a =
47$ meV for the LUMO orbital of pentacene.\cite{Cornil} 

We note that this band structure is similar to that for
the $\kappa$-(BEDT-TTF)$_2$X family\cite{McKenzie}.
A similarity between the band structure of
$\alpha$-sexithiophene and this family has been 
pointed out previously.\cite{Haddon}

\begin{figure}[tbp]
\vspace*{5cm}
\includegraphics{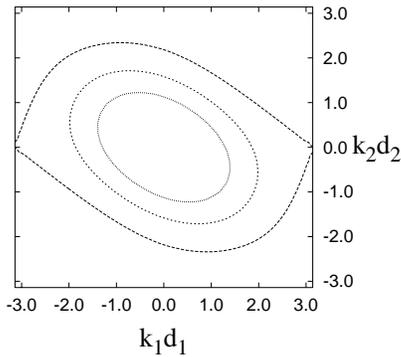}
\caption{Fermi surfaces for the band structure given in equation
(\ref{bandstru}), for different band fillings, $1/8$, $1/4$, $1/2$.}
\label{fermisurface}
\end{figure}  

Importantly, the coupling between the layers has been found to be
negligible, thus showing the strongly two-dimensional character of
these materials.\cite{Cornil}.  There are two molecules per unit cell
but the band structure reduces to one band due to the equivalence of
the crystal environment up to a rotation.

Correlation effects are known to be important for single
isolated acene molecules, as has been discussed in
the quantum chemistry literature\cite{Ramasesha,qchem}.
For the purpose of estimating the Hubbard repulsion, we have
calculated the H\"{u}ckel energies and wave-functions of the
electrons in benzene, naphthalene, anthracene, tetracene and pentacene
molecules.  We have then estimated the cost in Coulomb energy of two
additional electrons (or holes) on the molecule,

\begin{equation}
U = \int d^3r_1 d^3r_2 \phi^*_{\uparrow}(\vec{r}_1) \phi_{\downarrow}(\vec{r}_1) \frac{e^2}{|\vec{r}_1-\vec{r}_2|} \phi^*_{\uparrow}(\vec{r}_2) \phi_{\downarrow }(\vec{r}_2)
\label{Coulomb}
\end{equation}
\noindent
where $\phi_{\sigma}(\vec{r})$ is the lowest unoccupied molecular orbital
(LUMO) of the molecule. Using the H\"{u}ckel approach, we found the
coefficients $c_i$ defined by:

\begin{equation}
| \phi_{\sigma} \rangle = \frac{1}{\sqrt{N}} \sum_i c_i |i,\sigma \rangle
\label{TB}
\end{equation}

\noindent
where $| i \rangle$ is the atomic orbital of the $i$th carbon atom.
Inserting this into (\ref{Coulomb}) leads us to estimate the following
expression for $U$:

\begin{equation}
U= \frac{U_0}{N^2} \sum_i |c_i|^4
\end{equation}
 
\noindent
where the expression is written as a function of the \textit{local}
Coulomb interaction on a single carbon atom $U_0 \equiv \langle i,
\uparrow; i, \downarrow |e^2/|r_1-r_2| |i, \uparrow; i, \downarrow
\rangle$.  We have calculated the H\"{u}ckel parameters $c_i$ for the
four isolated molecules.  For the benzene molecule, the electron is
uniformly delocalized on the ring. The effective $U$ is therefore
$U_0/N$. For the other molecules, there is incomplete delocalization
and $\sum_i |c_i|^4 > \sum_i |c_i|^2 = N$. Therefore $U>U_0/N$. The
explicit calculation gives $U/U_0=0.14$ for naphthalene ($N=10$),
$0.12$ for anthracene ($N=14$), $0.10$ for tetracene ($N=18$) and
$0.09$ for pentacene ($N=22$). Assuming a value of $U_0=12$ eV for the
Coulomb interaction on a carbon atom,\cite{Ramasesha} one can estimate
that $U=1.7$ eV for anthracene, $U=1.2$ eV for tetracene and $U=1$ eV
for pentacene. Although this calculation neglects the screening
effects due to the other electrons, it suggests that the Coulomb
interaction ($U \sim 1$ eV) may be comparable to the kinetic energy (with
a bandwidth given by $W \sim 0.5 $ eV\cite{Cornil}). This is
consistent with other similar organic materials based, for example, on
the BEDT-TTF or BETS molecules.\cite{McKenzie} 
Although our  estimate of $U$ is rather crude (especially because it
neglects screening effects), it seems plausible to view a single
layer in isolation as a strongly correlated system.

We first review the physics of the one-layer model at
half-filling\cite{McKenzie} and low temperatures before considering
the effects due to the next layers.  At small $U \ll t$, the single
layer system, described by the Hamiltonian (\ref{hamiltonian}), is
metallic, because the frustration of the lattice gives an imperfect
nesting of the Fermi surface (see Fig. \ref{fermisurface}).  At large
$U \gg t$, the system is a Mott insulator and has a charge gap $U$.
The interaction between the spins defines a smaller energy scale,
$\sim t^2/U$.  In the intermediate regime of $U \sim t$, the question
of the nature of the ground state remains to be settled.  As
previously suggested on theoretical grounds,\cite{Moriya} a
superconducting phase may appear near the metal-Mott insulator
boundary. This may explain the phase diagram of half-filled organic
materials.\cite{Moriya,McKenzie} As the model (\ref{hamiltonian}) is
essentially the same as the one which describes
$\kappa$-(BEDT-TTF)$_2$X,\cite{McKenzie} it is tempting to suggest
that superconductivity may appear in acene field-effect
transistors, at half-filling.  In such an approach, the $T_c$ is
expected to increase with the Coulomb interaction.\cite{Moriya2}
Meanwhile, the larger the molecule, the smaller the $U$ is. Therefore,
a larger $T_c$ is expected for smaller molecules such as tetracene and
anthracene (if screening effects can be ignored).  However, the same
trend is found within the electron-phonon mechanism.\cite{Kato} If
such strongly correlated superconducting phases appear, Mott
insulating states should also appear in other acenes with a larger
$U/t$.  This could be achieved in principle in materials based on a
smaller acene molecule (to increase $U$), such as
naphthalene. Alternatively, $t$ might be decreased by pushing the
molecules further apart by intercalation or attaching side groups to
the molecules.\cite{Haddon2}

\section{A model for layered field-effect transistors}
\label{A model}

We now examine how the single layer approach is modified when one
takes into account the existence of the next layers. We have discussed
above the example of acene-based materials to give an idea of the order of
magnitude of the parameters, but the following considerations do not
actually depend on the details of these materials.

Our aim is to study the case $U \gg t$ to see whether the Mott state
we have discussed for a single layer can really be induced when the
system has many layers. As it turns out to be the case, we may then
speculate that superconductivity in field-effect transistors
would probably be of the same origin as that observed in bulk crystals
of $\kappa$-(BEDT-TTF)$_2$X. Conversely, the existence of
superconductivity in those materials should encourage the search for
superconductivity in field-effect transistors.

According to the band structure
calculations,\cite{MacDonald,Poilblanc} the electrons are confined
to a single layer for the largest electric fields applied.
Nevertheless, we do not restrict ourselves to a single band for two
reasons. First, even when the charges are confined to a single
layer, the next layers offer virtual states that make the system
different from bulk crystals. This is precisely what we study in the
next paragraph, where the interlayer coupling is set to zero, so that
there is no mixing of the bands.  Second, when the electric field is
small enough with respect to the interlayer coupling $t_{\perp}$, the
charges are partly delocalized onto the next layers. So the question
is then: can we still have a Mott insulating state in the first layer?

We now take into account the next layers away from the interface.  The
Hamiltonian includes the kinetic energy of the electrons
within a layer, the strong interaction between them, the electric
field imposed by the gate (but we neglect the modification of the
electronic orbital by the electric field) and the interlayer hopping:

\begin{eqnarray}
{\cal H} &=& \sum_{{\bf k},i,\sigma} \epsilon_k c^{\dagger}_{{\bf
k},i, \sigma}c_{{\bf k},i, \sigma} +  U \sum_{{\bf{i}},j} n_{{\bf{i}},j,\uparrow}
n_{{\bf{i}},j,\downarrow} \nonumber \\
 &+&  \Delta \sum_{{\bf k},i,\sigma} i \hspace{.1cm} n_{{\bf k},i, \sigma} - t_{\perp}  \sum_{{\bf k},i,\sigma} (c^{\dagger}_{{\bf
k},i,\sigma}c_{{\bf k},i+1, \sigma} + \mbox{h.c.})
\label{Hamiltonian}
\end{eqnarray}
where $c^{\dagger}_{{\bf k},i, \sigma}$ creates an electron in a state
of momentum ${\bf k}$, in the layer $i$ and spin state
$\sigma$. $\epsilon_k$ is the electronic dispersion within a layer,
characterized by a bandwidth $W$. $\Delta=eEc$ is the energy spacing
between the orbitals which belong to different layers, due to the
electric field $E$ imposed by the gate. $c$ is the distance between
the layers. We also define $\gamma=2\Delta/W$. The Hamiltonian is then
a multi-band Hubbard model with the set of parameters: $U/W$,
$\gamma$, $t_{\perp}/W$ and the total filling $n$.

\begin{figure}[tbp]
\centerline{
\psfig{file=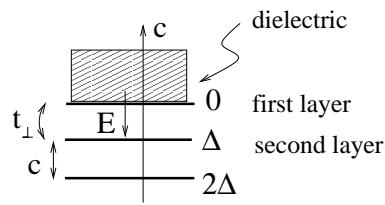,width=5cm} }
\vspace{0.2cm}
\caption{Schematic picture for the first layers near the interface
with the dielectric crystal. $\Delta=eEc$ is the energy associated
with the electric field.}
\label{layers}
\end{figure}

\subsection{No interlayer hopping}
\label{mottsection}

Firstly, we will consider the case of decoupled layers, $t_{\perp}=0$,
which is in fact a good approximation for acene materials.  The only
additional ingredient compared to section \ref{A strongly} is the
existence of additional states in the next layers at energies
$[\Delta_i-W/2,\Delta_i+W/2]$ (where $\Delta_i=i \Delta$), which
possibly overlap with the first band.

\begin{figure}[tbp]
\centerline{
\psfig{file=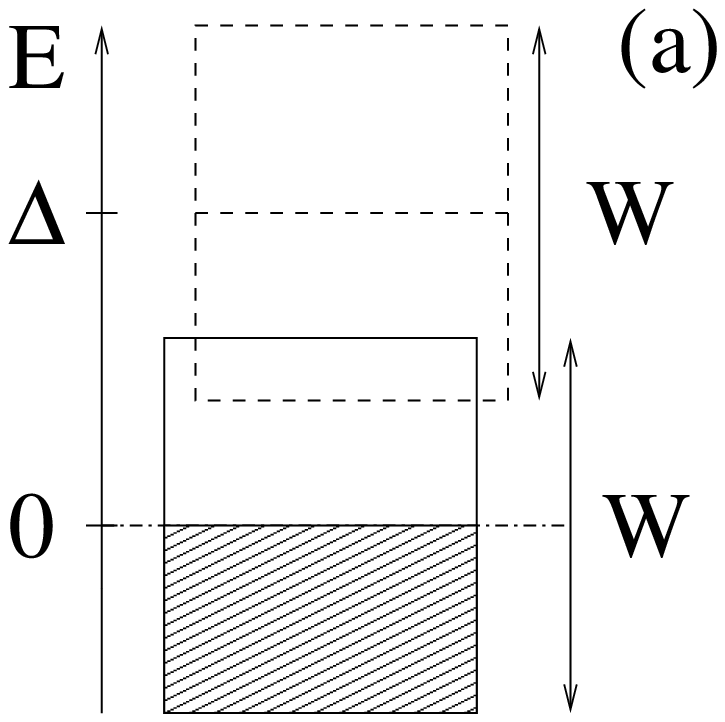,width=4.2cm} 
\psfig{file=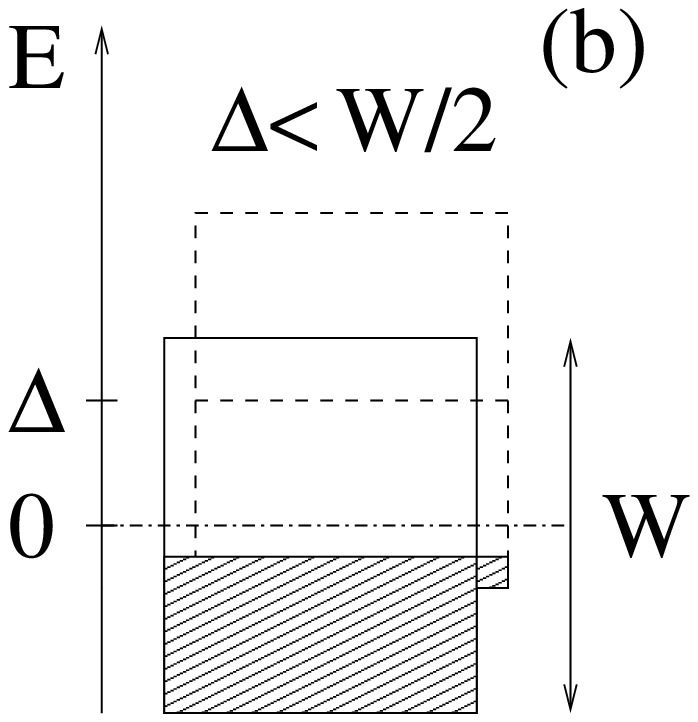,width=4.2cm}}
\vspace{0.2cm}
\caption{Non-interacting band picture with the first two bands for a
total filling of one electron per molecule, i.e., a total of $N$
carriers where $N$ is the number of sites within a layer, $W$ the
bandwidth and $\Delta=eEc$ the mid-energy of the conduction states of
the second layer. (a) $\Delta>W/2$. The lower band is
half-filled. (b) $\Delta<W/2$. The lower band is no longer half-filled
: some electrons occupy the states of the second band.}
\label{noninteracting}
\end{figure}

We now discuss the effects of these additional states for both the
case of half-filling and arbitrary fillings.

\medskip

\textit{Half-filling}. As long as $\Delta>W/2$, no additional states
are filled (Fig. \ref{noninteracting}(a)) and due to the absence of
coupling between the bands, the system behaves as a single layer in
isolation, i.e., there is a metal-insulator transition at finite $U$,
as we explained above.  The only difference with the single layer
problem is the existence of additional excited states, which will
affect the transport and magnetic properties of the system.  Thus,
depending on the strength of $U/\Delta$, the system is either a Mott
insulator ($U/\Delta<1$, the charge gap is roughly given by $U$) or a
charge-transfer insulator ($U/\Delta>1$), for which the excited states
consist of a charge transfer to the next layer (the charge gap is
given by $\Delta$). Thus, the gap can be much smaller than $U$ if the
electric field is not large enough. The magnetic properties of the
layer are given by superexchange processes. In the case of the Mott
insulator, the effective antiferromagnetic exchange is given as usual
by $J=4t^2/U$. In the case of the charge-transfer insulator, a given
electron would not necessarily hop virtually to a neighboring site in
the same layer because it costs the energy $U$ but would rather hop
onto the next layer which costs a smaller energy $\Delta=eEc$. Then,
in this case, the exchange $J$ between two spins would explicitly
depend upon the strength of the electric field.  Note, however, that
if we change $E$, we would not only change $J$, but also the total
filling of the system. Above half-filling, some carriers will occupy
states in the next layer. As long as the layers are weakly coupled,
these extra carriers will not change the magnetic properties of the
first layer. Therefore, the net effect of the electric field will be
to decrease the interaction between the spins.  This effect can not be
detected through the change of the critical temperature of the
antiferromagnetic transition because such transition is prevented by
the Mermin-Wagner theorem in a strictly two-dimensional
system. However, measurement of the magnetic susceptibility would
provide information about spin-spin interaction and how they change
with the electric field.  When $\Delta < W/2$ (see
Fig. \ref{noninteracting}(b)), however, the second layer starts to be
filled before the first one gets to half-filling (the chemical
potential is negative).  The entire system is less than half-filled
and no metal-insulator transition is then expected.
%The situation is
%summarized in Fig. \ref{half-filling}. 
The critical value $\Delta=W/2$ is very schematic here because
$t_{\perp}=0$. In the section \ref{effectof}, we will see that for
finite $t_{\perp}$, the critical value of $\Delta$ actually increases.

%\begin{figure}[tbp]
%\centerline{
%\psfig{file=phasediagram.ps,width=7cm}}
%\vspace{0.2cm}
%\caption{Schematic phase diagram for $t_{\perp}=0$ and $N$ carriers
%(so that $n_{2D}=1/2$). $\gamma>1$: the electrons are confined onto
%the first layer, because of the large electric field. $\gamma=
%2\Delta/W <1$: some states in the layers away from the surface are
%filled. The electronic density of the entire system is smaller than
%$1/2$. Therefore, the system is metallic. $U_M$ is the critical value
%for the Mott transition in a single layer.}
%\label{half-filling}
%\end{figure}  

\textit{Arbitrary fillings}. In the case $\Delta<W/2$, some states in
the second band have a negative energy and need to be filled before
the first band becomes half-filled (see Fig. \ref{noninteracting}(b)),
thus possibly allowing for a localization of the charges in the first
layer if $U$ is large enough. To assure that the first band is
half-filled, we fix the chemical potential at zero energy, $\mu=0$.
In the next paragraph, we calculate, in the framework of the
slave-boson theory, the total number of carriers needed to have
$\mu=0$.

In the slave-boson mean-field theory,\cite{Kotliar} the effect of the
Coulomb interaction is to renormalize the hoppings by quantities
denoted by $z_j^2$ (for the layer $j$). $z_j$ is the mean field value
of the operator $\hat{z}_{\textbf{i},j}$ which is given by a
combination of auxiliary boson operators,
$e_{\textbf{i},j}^{\dagger}p_{\textbf{i},j,\sigma}+p_{\textbf{i},j,-\sigma}^{\dagger}
d_{\textbf{i},j}$, that tell if the site $(\textbf{i},j)$ is empty
(operator $e_{\textbf{i},j}$), single occupied
($p_{\textbf{i},j,\sigma}$) or double occupied
($d_{\textbf{i},j}$).\cite{Fulde} The many-body Hamiltonian
(\ref{Hamiltonian}) for the layer $j$ reduces in the mean-field approximation to free fermions with
renormalized bandwidths:

\begin{equation}
{\cal H}_j = -\sum_{{\bf{il}},\sigma} t_{\bf il}  z_j^2 (c^{\dagger}_{{\bf{i}},j,\sigma}
c_{{\bf{l}},j,\sigma} + \mbox{h.c.})  +  U d_j^2  
 +  \Delta \sum_{{\bf i},\sigma} j \hspace{.1cm} n_{{\bf i},j, \sigma}
\label{SBHamiltonian}
\end{equation}
where $d_j$ is the mean-field value of the double occupancy
operator $d_{\textbf{i},j}$. In the following, we will restrict ourselves to $U
\rightarrow +\infty$, so that $d_j=0$ (the double occupancies are
forbidden).  The layers are all decoupled and the total Hamiltonian is
the sum over all the layers. However, the quantities $z_j^2$, which
determines the bandwidths, are coupled together because they depend on
the fillings of the band $j$, which is determined by the chemical
potential of the entire system.

For free electrons ($U=0$, $z_j^2=1$), the number of states below zero
energy in the second layer is denoted by $(1-\gamma)N$ ($\gamma<1$ and
$N$ is the number of sites).  When we take into account the Coulomb
interaction within the slave-boson picture, the bandwidths are reduced
($W \rightarrow Wz_2^2$). The number of states given above becomes
$(1-\gamma/z_2^2)N$, that is smaller than $(1-\gamma)N$ and depends
upon $z^2_2$ that we have to determine. Therefore, we note that the
total filling needed to have $\mu=0$ in the interacting system must be
smaller than that for free electrons. We now calculate the filling
explicitly.

We first calculate the value of $z_2^2$ for $\gamma<1$ when the
carriers of the first layer become localized. For that, we consider
just the first two layers ($j=0,1$) and take a \textit{constant}
density of states in the layers to simplify the calculation. At
$\mu=0$, the fillings $n_j$ are given by:

\begin{eqnarray}
n_1 &=& \frac{1}{2}+ \frac{(n-1)z^2_2+\gamma/2}{z_1^2+z_2^2} \label{dens1} \\
n_2 &=& n-n_1
\end{eqnarray}
where $n$ is the total filling of the system, $Wz_1^2$ and $Wz^2_2$
are the bandwidths of the two bands.  In the limit of large $U$,
(i.e., $U \gg W, \Delta$),
$z^2_1=(1-2n_1)/(1-n_1)$. Charge localization in the first layer is
characterized by $z^2_1=0$, or $n_1=1/2$. First, the equation
(\ref{dens1}) gives the bandwidth of the second band
$z_2^2=\gamma/[2(n-1)]$.  On the other hand, in the limit of large
$U$, the bandwidth is given by $z^2_2=(1-2n_2)/(1-n_2)$, where
$n_2=n-1/2$. Equating the two expressions, we find the total filling
$n$, $n_c(\gamma)$, for which $n_1=1/2$ and the bandwidth $z_2^2$:

\begin{eqnarray}
z^2_2 &=& \frac{1}{2}(\sqrt{\gamma^2+8\gamma}-\gamma) \label{renorm} \\ n_c(\gamma)
&=& 1-\frac{1}{\sqrt{1+8/\gamma}-1} \equiv 1-f(\gamma)
\label{truc}
\end{eqnarray}
Note that this solution is valid as long as the third band is not
involved, that is to say, if we define the energy of the middle of the
third band by $\Delta_2=\alpha_2 \Delta$, $\Delta_2-W/2>0$, or in
terms of $\gamma$, $\gamma > 1/\alpha_2$. 

\begin{figure}[tbp]
\centerline{
\psfig{file=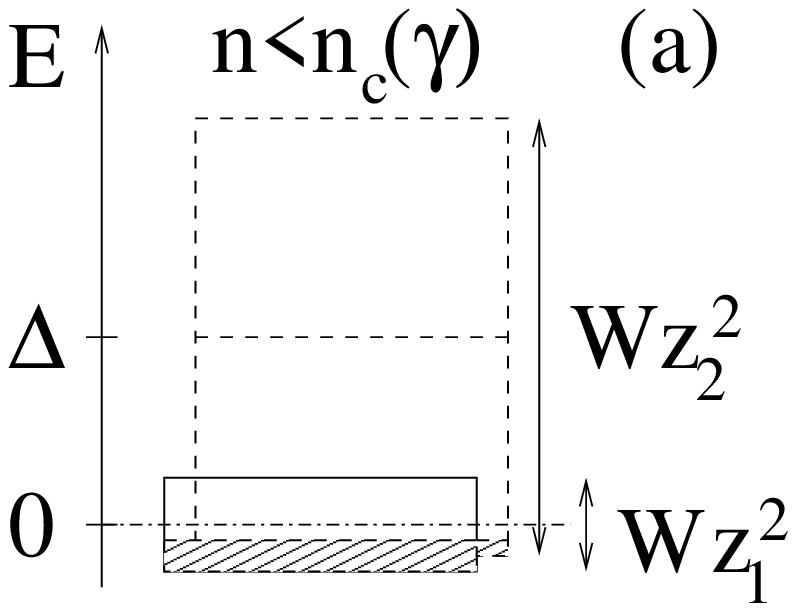,width=4.2cm}
\psfig{file=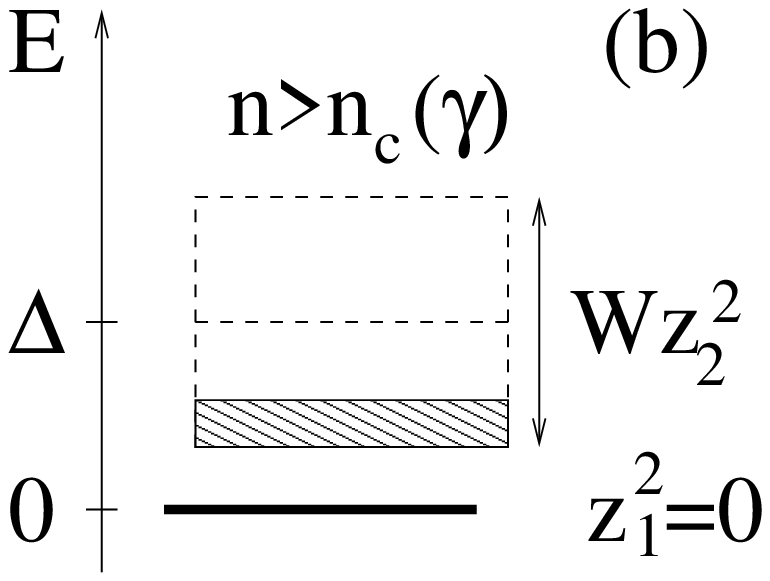,width=4.2cm}}
\vspace{0.2cm}
\caption{Interacting band picture for the first two bands within the
slave-boson mean-field theory (here $2\Delta<W \ll U$). (a) The first two
bands are partially filled with electrons and the bandwidths are
renormalized by the interaction. (b) As soon as $n>n_c(\gamma)$, the
lower band is half-filled, so that $z^2_1=0$. Then, there is one
electron localized on each site of the first layer. The second band is
partially filled.}
\label{fig5b}
\end{figure}

Therefore, in the regime $ 1/\alpha_2 <\gamma <1$, the total filling
needed to get the first band half-filled is given by
eq. (\ref{truc}). Above this filling, $z_1^2=0$, so that we have one
localized electron on each site of the first layer. The additional
$(2n_c(\gamma)-1)N$ electrons fill the band of the second layer which
bandwidth is $Wz_2^2$ (given by eq. (\ref{renorm}), see
Fig. \ref{fig5b}).

More generally, in order to consider the regime with $\gamma <
1/\alpha_{N_b}$, we have to consider $N_b$ additional bands.
Generalization to many bands with energies $\Delta_i=\alpha_i \Delta$
($\alpha_0=0$, $\alpha_1=1$, $\alpha_2$...) and bandwidth $W$ is
straightforward. Performing the same steps as above, in the regime
$1/\alpha_{N_b+1}<\gamma<1/\alpha_{N_b}$, the filling at which the
$N_b$ first bands are half-filled is given by:

\begin{equation}
n_c(\gamma)=\frac{N_b+1}{2} - \sum_{p=1}^{N_b} f(\alpha_p \gamma)
\label{P=1}
\end{equation}
Let us recall that $N_b$ is the number of bands partially filled. For
a given system with a fixed $\gamma$, $N_b$ bands are involved when the
bottom of the $N_b^{th}$ band crosses the zero energy. It occurs when
$\gamma$ gets smaller than $1/\alpha_{N_b}$.The shape of this curve
depends on the positions of the bands, $\Delta_P=\alpha_P \Delta$.

\begin{figure}[htbp]
\centerline{
\psfig{file=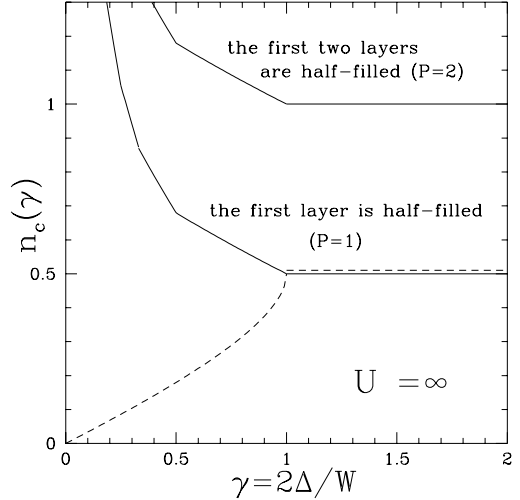,width=7cm}}
\vspace{0.2cm}
\caption{Phase diagram (filling, $n$, versus energy spacing between
the bands due to the electric field, $\gamma$) from slave-boson theory
in the limit $U=\infty$ and $t_{\perp}=0$. Solid lines show the
critical fillings above which the first (resp. second) band is
half-filled, i.e. the electrons are localized on each site of the
first (resp. second) layer in the limit of large $U$. The dashed line
shows the critical filling above which the second band starts to be
filled.}
\label{phaseD}
\end{figure}

Furthermore, the generalization to higher fillings is also very
similar. If we consider such a filling that $P$ bands are completely
half-filled and $N_b$ bands are partially filled. In the regime where
$1/(\alpha_{P+N_b}-\alpha_{P-1})<\gamma<1/(\alpha_{P+N_b+1}-\alpha_{P-1})$,
the critical $n_c$ to get the $P^{th}$ band half-filled is given by:

\begin{equation}
n_c(\gamma)=\frac{N_b+P}{2} - \sum_{j=P}^{N_b+P-1} f[(\alpha_j-\alpha_{P-1}) \gamma]
\label{Pqcq}
\end{equation}
Note that this equation reduces to (\ref{P=1}) for $P=1$.  Examples of
the first two curves ($P=1$ and $P=2$) are given in figure \ref{phaseD}.

In conclusion, when $\gamma>1$, only one band is involved. When the
band gets to half-filling ($n=1/2$), it undergoes a transition to a
Mott state at large $U$ in which all the electrons are localized on
the first layer. Doping the system further leads to filling of the
states of the second layer, eventually up to half-filling when the
total density is $n=1$, then localizing the electrons in the second
layer, and so on (see Fig. \ref{phaseD}). However, as soon as
$\gamma<1$ (the second band starts to be involved above some filling, see dashed line in Fig. \ref{phaseD}),
the Mott state needs a larger filling to occur.  This filling,
$n_c(\gamma)$, is given by eq. (\ref{P=1}). In that case, the first
band is half-filled, and there is one electron localized on each
site. As the total number of electrons is larger than the number of
sites of the first layer, some of them occupy the states of the next
layers.  The system is metallic, but with a reduced number of
carriers.

The condition on $\gamma$ is nothing but a condition on the electric
field since $\gamma=2eEc/W$. The border $\gamma=1$ defines a critical
field $E_c$.  For instance, in the acenes, $W \sim 0.2$ eV and $c \sim
15 \rm \AA $ gives $E_c \sim 10^8 $ Vm$^{-1}$, which is close to the
dielectric breakdown field.\cite{dielectricconstant}

However, this calculation neglects the interlayer hopping. It is
known that such couplings are quite small in layered molecular
crystals, in particular in those discussed above,\cite{Cornil,Cornil2,
Haddon,Haddon2} but it is
not obvious whether an infinitesimal coupling would destroy the charge localization in either layers. In the next section, we study whether such a state
survives when one takes into account such interactions, such as the
interlayer coupling or also the Coulomb interaction between nearest
sites.

\subsection{Effect of the interlayer hopping}
\label{effectof}

The Mott state in a single layer of a field-effect transistor is not
as stable as in a bulk crystal. The electrons have, indeed, other
available conduction states in the next layers away from 
the surface. This opens a way to avoid the strong on-site Coulomb
interaction, though it costs the energy associated with the electric
field, $eEc$.  Some interactions will actually act against the charge localization: the coupling between the layers favors extended states away
from the surface. We just consider the hopping between the two layers closest to the surface:

\begin{equation}
{\cal H}_{\perp} = - t_{\perp}  \sum_{{\bf k},\sigma} (c^{\dagger}_{{\bf
k},1,\sigma}c_{{\bf k},2, \sigma} + \mbox{h.c.})
\end{equation}
When $\gamma>1$, the two bands are separated by a gap.  Using the same
Hamiltonian, but at half-filling, Blawid \textit{et al.} showed that,
within the slave-boson theory, there is a Mott-metal transition when
$t_{\perp}>W[\gamma(\gamma-1)/2]^{1/2}/2 \equiv t_{\perp c} $ for
$U=+\infty$.\cite{Fulde1} Note that, when $\gamma \rightarrow 1$, the
critical $t_{\perp}$ vanishes. This is consistent with the
disappearance of the Mott state at half-filling that we found when
$\gamma<1$.

On the other hand, in section \ref{mottsection}, we found that a total
filling larger than $n_c(\gamma)$ ($>1/2$) was needed, when
$\gamma<1$, to have the first layer half-filled, thus allowing for a
Mott state in the first layer. We now wonder if this state is still
stable above $n_c(\gamma)$ when $t_{\perp} \neq 0$. If we think of the
result given above, it seems that we need a real gap between the two
bands ($\gamma > 1$). With the Coulomb interaction switched on, the
bandwidth of the second band is reduced and a gap opens when
$\Delta-Wz_2^2/2 > 0$.  In the limit of large $U$, $z_2^2$ is given by
$(1-2n_2)/(1-n_2)$, and $n_2=n-1/2$ as long as any other band can be
safely ignored.  Therefore, the gap opens when the density $n$
satisfies $z_2^2=\gamma=(2-2n)/(3/2-n)$. That gives
$n=(2-3\gamma/2)/(2-\gamma) \equiv n_c^{\prime}(\gamma) >
n_c(\gamma)$. So we could think that the Mott state is unstable in the
regime $n_c(\gamma)<n<n_c^{\prime}(\gamma)$, but becomes stable as
soon as a gap opens, i.e. when $n>n_c^{\prime}(\gamma)$.

In fact, the case considered here is different from that of Blawid
\textit{et al.} because the second band is partially filled. We now
show that, as soon as $n>n_c(\gamma)$, the Mott state is stable below
a critical value for $t_{\perp}$. This is because we do not need a gap
between the two bands, but a gap between the energy of the localized
states and the energy of the available states in the second band. As
soon as $\mu>0$, the second band is partially filled and the gap
opens.

We consider \textit{two} coupled bands with dispersions given by
$\epsilon_k$ and $\tilde{\epsilon}_k$ ($\tilde{\epsilon}_k=\Delta +
\epsilon_k $ in the non-interacting picture). We neglect the other
higher-energy bands in order to simplify the expressions.  In the
slave-boson picture, the effect of the interaction is just to
renormalize the bandwidths by quantities denoted by $z_1^2$ and
$z_2^2$.

\begin{eqnarray}
{\cal H}_0 &=& \sum_{{\bf k},\sigma} z_1^2 \epsilon_k c^{\dagger}_{{\bf
k}, 1, \sigma}c_{{\bf k}, 1, \sigma} + \sum_{{\bf k},\sigma} \underbrace{(\Delta + z_2^2 \epsilon_k)}_{\tilde{\epsilon}_k} c^{\dagger}_{{\bf
k}, 2, \sigma}c_{{\bf k}, 2, \sigma}  
\end{eqnarray}
It also renormalizes the coupling
between the two bands such as:

\begin{equation}
{\cal H}_{\perp} = - t_{\perp} z_1z_2 \sum_{{\bf k},i,\sigma}
(c^{\dagger}_{{\bf k},1,\sigma}c_{{\bf k},2, \sigma} + \mbox{h.c})
\end{equation}
We again restrict ourselves to the limit of infinite $U$, where the
$z_i^2$ are function of the fillings $n_i$:

\begin{eqnarray}
z_i^2 &=& \frac{1-2n_i}{1-n_i} 
\label{zcarre}
\\ n_i &=& \frac{1}{2N} \sum_{k \sigma} \langle
c_{k,i,\sigma}^{\dagger} c_{k,i,\sigma} \rangle 
\end{eqnarray}
where the $n_i$ are calculated as function of the parameters of the
model. Near the metal-insulator boundary for the first layer, (but in
the metallic region), $n_1$ will be expanded as a function of $z_1^2 \rightarrow 0$:

\begin{equation}
n_1=\frac{1}{2}-\beta z_1^2
\label{n1}
\end{equation} 
where we have introduced a parameter $\beta$ that has to be
determined.  To satisfy equations (\ref{zcarre}) and (\ref{n1}) at
the same time, in the metallic regime, we need:

\begin{equation}
4 \beta = 1
\label{critic}
\end{equation}
This gives a condition on the parameters of the model at the Mott-metal boundary. So all we have
to do is to calculate this parameter $\beta$.  We decompose $c_{k, 1,
\sigma}^{\dagger}$ as function of the eigen-operators:

\begin{equation}
c_{k, 1, \sigma}^{\dagger}=u_k \alpha_{k  \sigma}^{\dagger} + v_k \beta_{k    \sigma}^{\dagger}
\end{equation}
where $u_k$ and $v_k$ are given by the diagonalisation of the
Hamiltonian.  We are interested in the limit $t_{\perp} \sim z_1
\rightarrow 0$. Then, these parameters are expanded as:

\begin{eqnarray}
|u_k|^2 & \sim &  1- \frac{t_{\perp}^2}{ \left( \epsilon_k - \tilde{\epsilon}_k \right)^2} \\
|v_k|^2 & \sim &  
\frac{t_{\perp}^2}{ \left( \epsilon_k - \tilde{\epsilon}_k \right)^2}
\end{eqnarray}
and $n_1$ can be rewritten:

\begin{equation}
n_1= \frac{1}{2N} \sum_k |u_k|^2 \langle \alpha_{k  \sigma}^{\dagger} \alpha_{k  \sigma} \rangle + \frac{1}{2N} \sum_k |v_k|^2 \langle \beta_{k  \sigma}^{\dagger} \beta_{k  \sigma} \rangle.
\end{equation}

\subsubsection{$\mu=0$} We restrict ourselves to $\gamma>1$, as the Mott state appears to be unstable when $\gamma=1$. Then only the lower band is filled, i.e. $ \langle \beta_{k  \sigma}^{\dagger} \beta_{k  \sigma} \rangle=0$ and $n_1$ is given by:

\begin{equation}
n_1 = \frac{1}{2} - \frac{1}{2N} \sum_k \frac{t_{\perp}^2}{ \left( \epsilon_k -
\tilde{\epsilon}_k \right)^2} 
\end{equation}
where the sum runs over all the $k$.  At the first order in $z_1^2$,
we can replace $\epsilon_k$ by $0$ in the expression
above. $\tilde{\epsilon_k}$ has a bandwidth which depends on
$z_2^2$. As the second band $\beta$ is empty, $n_2$ is of the order
${\cal O}(t_{\perp}^2)$. Then, using (\ref{zcarre}), $z_2^2$ is just
$1-{\cal O}(t_{\perp}^2)$.  We can replace $\tilde{\epsilon}_k -
\epsilon_k $ by $\Delta + \epsilon_k$:

\begin{equation}
n_1 = \frac{1}{2} - \frac{1}{2N} \sum_k \frac{t_{\perp}^2}{(\Delta +
\epsilon_k)^2}
\end{equation}
which can be expressed as a function of an integral:

\begin{equation}
n_1 = \frac{1}{2} - \frac{1}{2} \int_{-W/2}^{0} d\omega \rho_0(\omega)
\frac{t_{\perp}^2}{(\Delta + \omega)^2}
\end{equation}
with $\rho_0=2/W$, the density of states per site that we have taken
constant. The upper limit is $0$, because only half of the first band
is filled. Now, we make the substitution $t_{\perp} \rightarrow
t_{\perp} z_1$:

\begin{equation}
n_1 = \frac{1}{2} - \frac{t_{\perp}^2 z_1^2}{2\Delta(\Delta-W/2)}
\end{equation}
therefore, the critical $t_{\perp}$, given by eq. (\ref{critic}), is:

\begin{equation}
t_{\perp c} =  \sqrt{\Delta(\Delta-W/2)/2}
\label{simple}
\end{equation}
which is the expression found previously.\cite{Fulde1} Along the line
$\gamma>1$, $n=1/2$, the Mott state is stable below a critical
$t_{\perp}$ which is given above. Note that it vanishes when the gap between the two bands vanishes, $\gamma
\rightarrow 1$ ($\Delta \rightarrow W/2$).

\medskip

\subsubsection{$\mu>0$} 
\begin{figure}[tbp]
\centerline{
\psfig{file=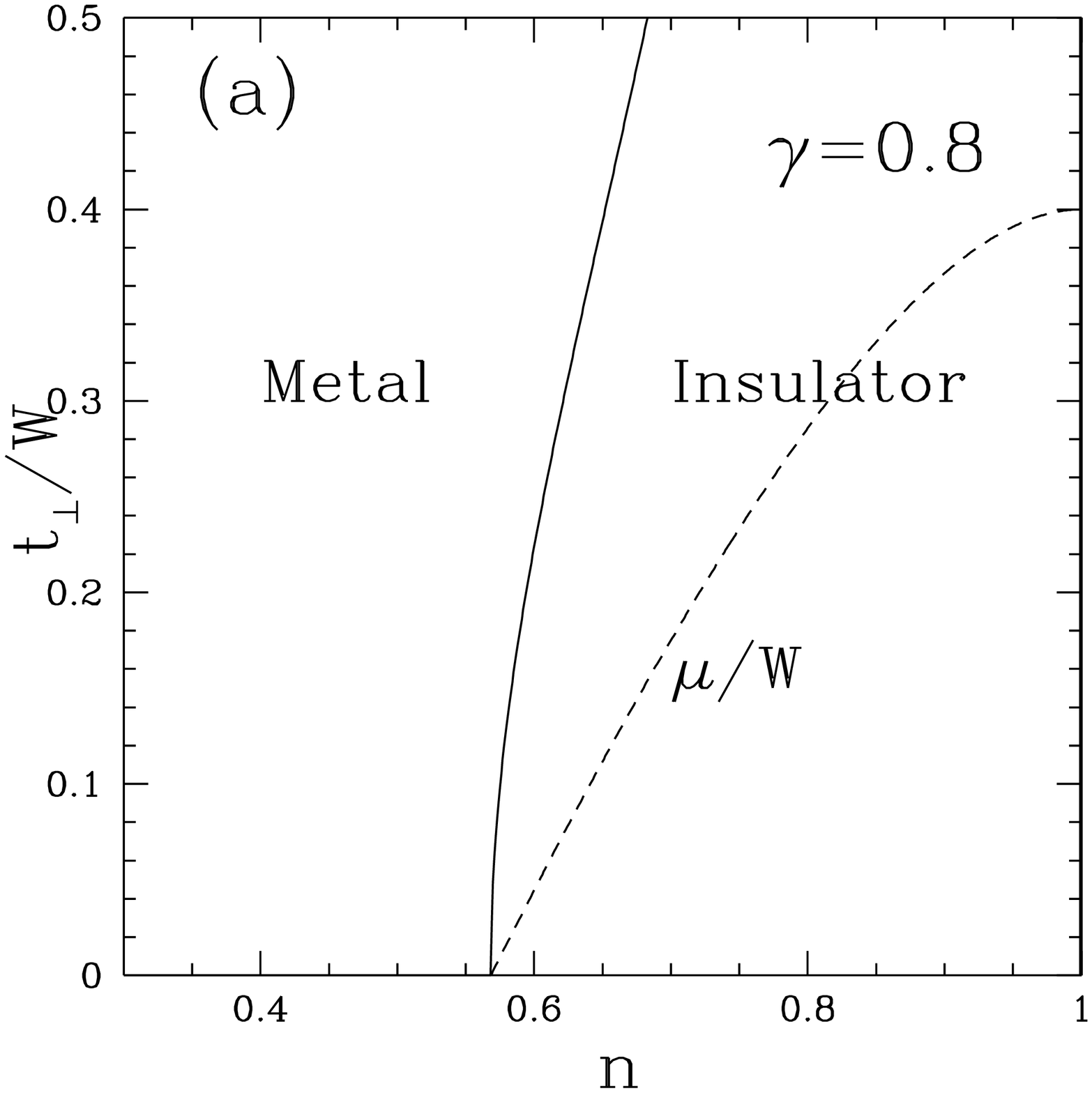,width=4.5cm} 
\psfig{file=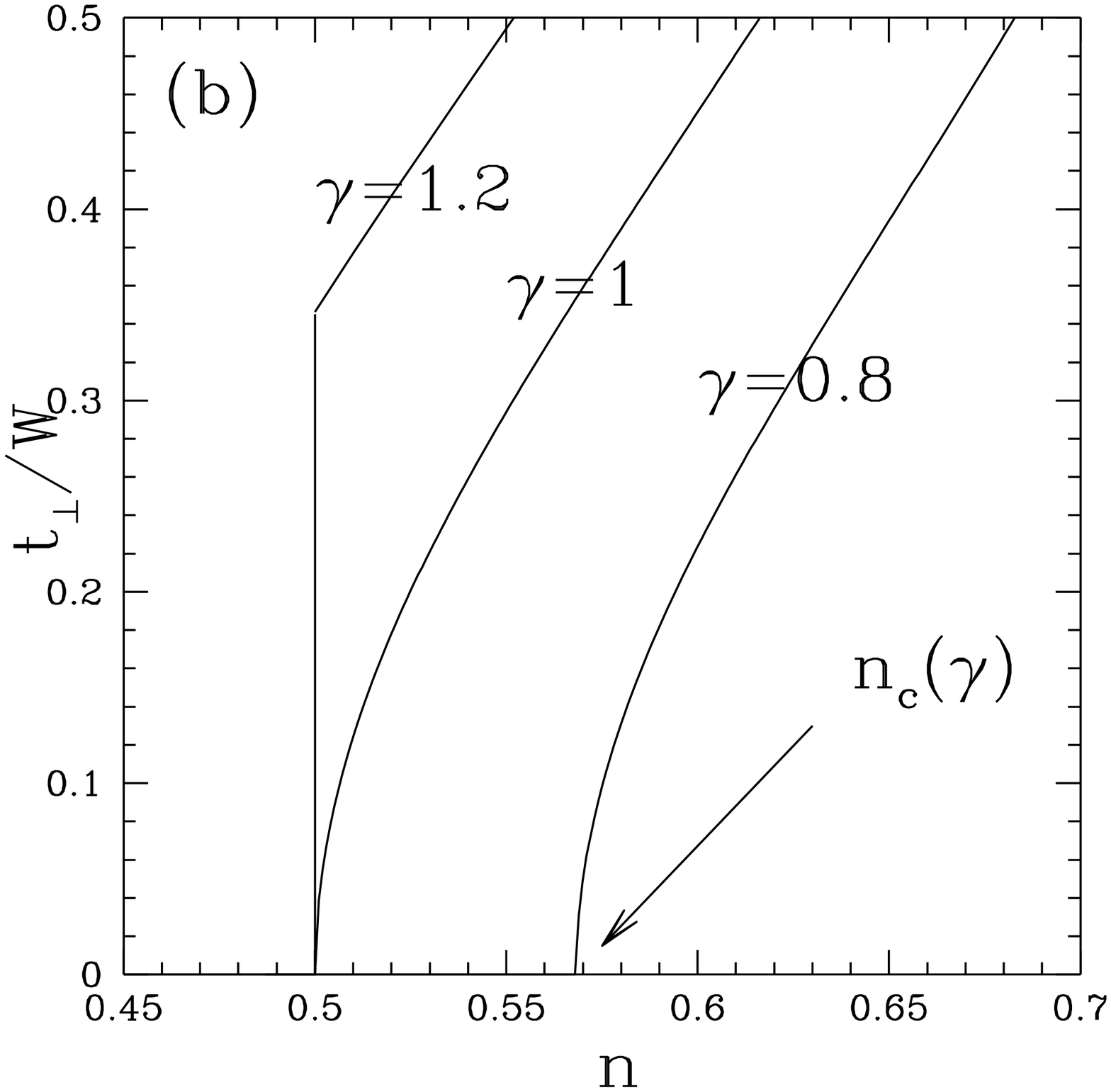,width=4.5cm}}
\vspace{0.2cm}
\caption{(a) Example of the phase diagram when $t_{\perp} \neq 0$, for
$\gamma=0.8$. Charge localization on each site of the first layer
occurs as soon as the chemical potential $\mu >0$, i.e. as soon as there is a gap between the
energy of the localized states and the energy of the available states
in the second band. (b) The critical line for different values of
$\gamma$. In every case, $U=\infty$.}
\label{tperpfig}
\end{figure}

We have to fill the
second band $\beta$ with electrons. $n_1$ has an additional term, and
is given by:

\begin{equation}
n_1 = \frac{1}{2} - \frac{1}{2N} \sum_k \frac{t_{\perp}^2}{
\tilde{\epsilon}_k^2} + \frac{1}{2N} \sum_k \frac{t_{\perp}^2}{
\tilde{\epsilon}_k^2} \langle \beta_{k  \sigma}^{\dagger} \beta_{k 
\sigma} \rangle
\end{equation}
where $\langle \beta_{k \sigma}^{\dagger} \beta_{k \sigma} \rangle=
\theta(\tilde{\epsilon}_k-\mu +{\cal O}(t_{\perp}^2))$. We have now to
calculate the third term. The second one is, indeed, identical,
provided that $W \rightarrow W z_2^2$ in order to take into account
the renormalized bandwidth of the second band due to its filling. We
first calculate the integral and then make the substitution $W
\rightarrow Wz_2^2$.

\begin{equation}
 \int_{-W/2}^{\mu-\Delta} d \omega \rho_0(\omega) \frac{t_{\perp}^2}{
(\Delta+\omega)^2} = \frac{t_{\perp}^2}{W}
\frac{\mu-\Delta+W/2}{\mu(\Delta-W/2)}
\end{equation}
Therefore, after the substitution, we find:

\begin{equation}
n_1 = \frac{1}{2} - \frac{t_{\perp}^2}{2\Delta(\Delta-Wz_2^2/2)}
+  \frac{t_{\perp}^2}{Wz^2_2} \frac{\mu-\Delta+Wz^2_2/2}{\mu(\Delta-Wz^2_2/2)}
\end{equation}
The next question is how $\mu$ is related to $n$.

\begin{equation}
n - n_1 = \frac{1}{2} \int_{-W/2}^{\mu-\Delta} d \omega \rho (\omega) 
\end{equation}
where $\rho(\omega)$ is the density of states of the second band,
which is affected by the coupling $t_{\perp}$. At the leading order,
however, we can replace $\rho(\omega)$ by $\rho_0(\omega)$ and $n_1$
by $1/2$. That gives the chemical potential:

\begin{equation}
\mu = \Delta+(n-1)W z_2^2 + {\cal O}(t_{\perp}^2)
\end{equation}
The problem is that $z_2^2$ is also function of the filling $n_2$. But
here again, because we are looking for the leading order in $z_1^2$,
we can replace $n_2$ by $n-1/2$. Then,

\begin{equation}
z_2^2 = \frac{1-2n_2}{1-n_2} = \frac{2-2n}{3/2-n}
\label{z2carre}
\end{equation}
Putting all the terms together and noticing that $t_{\perp}$ scales as
$z_1z_2$, $n_1$ is then given by:

\begin{equation}
n_1 = \frac{1}{2} - \left( \frac{t_{\perp}z_1}{W} \right)^2
\frac{\frac{4-4n}{3/2-n}}{\gamma-\frac{2-2n}{3/2-n}} \left(
\frac{1}{\gamma} - \frac{2n-1}{\gamma+2(n-1)\frac{2-2n}{3/2-n}}\right)
\nonumber
\end{equation}
and the equation for the stability border, $4\beta_c=1$, is written as:

\begin{equation}
\frac{t_{\perp c}}{W} = \left[  \frac{\frac{4-4n}{3/2-n}}{\gamma-\frac{2-2n}{3/2-n}}  \left( \frac{1}{\gamma} - \frac{2n-1}{\gamma+2(n-1)\frac{2-2n}{3/2-n}}\right)\right]^{-1/2} 
\end{equation}
The resultant phase diagram is given in Fig. \ref{tperpfig}(a) for a
particular value of $\gamma<1$. Below the critical $t_{\perp c}$, the
first layer is insulating. The $t_{\perp c}$ vanishes at a particular
density.  As shown in Fig. \ref{tperpfig}(a), it vanishes with $\mu$ ,
which corresponds to the density $n=n_c(\delta)$ that we calculated in
the previous section (for the present problem of two bands, see
eq. (\ref{truc}) or eq. (\ref{P=1}) or (\ref{Pqcq}) when more bands
are involved). In other words, $t_{\perp c}$ vanishes when there is no
gap between the energy of the localized states and the energy of the
available states in the second band.  In conclusion, as soon as
$n>n_c(\delta)$, there is one electron localized on each site of the
first layer for $t<t_{\perp c}$.  Remember, however, that we are
considering the $U=\infty$ limit.

Note that, when $n=1/2$, the second term in the brackets vanishes,
leading to the expression (\ref{simple}) which is valid for
$\Delta>W/2$ ($\gamma>1$).  

In Fig. \ref{tperpfig}(b), the critical $t_{\perp c}$ is given for
several values of $\gamma$. When $\gamma>1$ (see the case
$\gamma=1.2$), there is a critical $t_{\perp}$ even at $n=1/2$. When
$n$ increases above $1/2$, the bandwidth of the second band decreases
according to eq. (\ref{z2carre}) and the gap increases. Consequently,
the mixing of the two bands decreases and we need a larger $t_{\perp
c}$ to destroy the Mott state.  At $n=1$, it diverges. In the model we
are considering, we have only two bands. When $n=1$, the entire system
is half-filled and is therefore a Mott insulator. The $t_{\perp c}$ to
destroy it is then infinite because we are in the limit of infinite
$U$.

\subsection{Inter-site Coulomb Interaction}
\label{Inter-site}

Another example of an interaction that can destroy the Mott state is
given by the Coulomb interaction between the nearest neighbor sites:

\begin{equation}
{\cal H} = V \sum_{<i,j>} n_in_j
\end{equation}
This term is diagonal in the basis of localized electrons. In the
limit of large $U$ and at half-filling, the energy per electron of the
Mott insulating state is $E_0=zV/2$, where $z$ is the coordination
number of the lattice. Therefore, we expect that if $E_0$ becomes
larger than the energy spacing between bands, $\Delta$, it will be
energetically favorable to transfer carriers from the Mott insulating
layer to the next band.  The system will then be metallic.  Therefore, in
the limit of infinite $U$, we need $V<2\Delta/z$ to get a Mott state.

\section{Conclusion}

We have given an estimate of the on-site Coulomb interaction for the
layered acene molecular crystals. It turns out that the interaction
($U \sim 1$ eV) may be comparable to the band width     ($W \sim
0.5$ eV).  However, whether the system is in the metallic or
insulating regime, has to be determined experimentally. A prediction
of the nature of the real ground state would indeed need a more
accurate estimation of $U/W$.  This is in fact a common feature of
many organic materials\cite{McKenzie}.
 
In bulk crystals, a large repulsive interaction leads to a Mott
insulating state when there is one electron per site. In such a case,
the hopping of the electrons to the other sites costs the large energy
$U$. In field-effect transistors, however, when the first molecular
layer at the interface reaches half-filling, the electrons still have
many empty states available on the next layers away from the surface,
though these states are at higher energies depending on the strength
of the electric field.  When the electric field is smaller than a
critical value, $E_c=W/(2ec)$, the carriers occupy the conduction
states of the next layer and the first layer is doped with holes. In
this case, the Mott state is destroyed. However, above the critical
electric field, the Mott state is stable at half-filling, at least
below a critical value for the interlayer coupling, $t_{\perp c}$
(given by eq. (\ref{simple})).

On the other hand, we have shown that it is possible to restore the
Mott state below $E_c$ provided that the system is further doped.
This is in order to fill the first band up to half-filling, to
compensate for the loss of carriers which go onto the next layers.  At
$n=n_c(\gamma)$, the chemical potential vanishes    and allows for a
Mott state in the first layer. This is what we have found in section
\ref{mottsection}, neglecting the coupling between the layers. This coupling,
although small in the layered molecular crystals, is still present. In
section \ref{effectof}, we have calculated the critical value,
$t_{\perp c}$, below which the Mott state is stable.  It turns out
that as soon as $n$ reaches $n_c(\gamma)$, $t_{\perp c}$ becomes non-zero.

In reality, in a field-effect transistor, the doping $n$ and the gate
voltage (or equivalently the electric field $E$) are not independent
quantities. On Fig \ref{tperpfig}(b), if we take a particular material
with a fixed $t_{\perp}$ and $W$, when the electric field increases,
the metal-Mott boundary will be crossed at some stage (provided that
$U$ is large enough).

Finally, we speculate about the possibility of superconductivity in
organic field-effect transistors. Since a Mott state can be induced in
the first layer if the interaction is large enough, it is possible
that superconductivity can occur due to the strong correlation between
the electrons (though not strong enough to create a Mott insulating
state). Is it possible to find a superconducting phase near the
metal-insulator boundaries as in $\kappa$-(BEDT-TTF)$_2$X? Then, if
the superconductivity is induced in the first layer, does it effect
the next layers by the proximity effect?  Experimentally, a strong
electric field ($E \sim W / (c e)$), is needed, either to reach a
sufficient doping or to get a sufficiently large energy separation
between the bands. It is not obvious whether the regime in question
can be experimentally reached or not because of dielectric breakdown;
but it turns out that layered organic molecular crystals are good
candidates to observe such effects because of their relatively small
bandwidth and weak interlayer coupling.

\acknowledgements This work was supported by the Australian Research
Council. We thank J.S. Brooks, A.R. Hamilton, and U. Lundin for helpful
discussions.

\end{document}